\documentstyle[aps,prl,multicol,epsf]{revtex}

\begin{document}
\draft
\title{Exactly solvable analogy of small-world networks}

\author{S.N. Dorogovtsev$^{1, 2, \ast}$ and J.F.F. Mendes$^{1,\dagger}$}

\address{
$^{1}$ Departamento de F\'\i sica and Centro de F\'\i sica do Porto, Faculdade de Ci\^encias, 
Universidade do Porto\\
Rua do Campo Alegre 687, 4169-007 Porto -- Portugal\\
$^{2}$ A.F. Ioffe Physico-Technical Institute, 194021 St. Petersburg, Russia 
}

\maketitle

\begin{abstract}
We present an exact description of a crossover between two different regimes of simple analogies of small-world networks. Each of the sites chosen with a probability $p$ from $n$ sites of an ordered system defined on a circle is connected to all other sites selected in such a way. Every link is of a unit length. Thus, while $p$ changes from 0 to 1, an averaged shortest distance between a pair of sites changes from $\bar{\ell} \sim n$ to $\bar{\ell} = 1$. We find the distribution of the shortest distances $P(\ell)$ and obtain a scaling form of $\bar{\ell}(p,n)$. In spite of the simplicity of the models under consideration, the results appear to be surprisingly close to those obtained numerically for usual small-world networks.
\end{abstract}


\begin{multicols}{2}

\narrowtext

A sharp interest on small-world networks began last year, when a beautiful  paper of Watts and Strogatz \cite{watts} was published. A number of publications on the topic is rapidly increasing  \cite{coll,her,pan,barth,barr1,mon,new1,barr2,men,kas,new2,kul,mou1,mou2,alb,barab}. One of versions of the model proposed by Watts and Strogatz looks as the following (see ref. \cite{new1}). Let $n$ sites  be arranged on a circle. Each of several neighbours are connected by unit length links. Then each of those links is rewired with a probability $p$ to one of other randomly chosen sites. Thus, one studies a crossover from an ordered structure at $p=0$ to a random graph \cite{ball}. 

The main quantity which is considered for these models is the shortest distance between two sites of a network (that is a quantity which is much more difficult to calculate than connectivity, for instance). When $p=0$, the averaged over all pairs of sites shortest distance $\bar{\ell}$ is equal to $n/4$, and $\bar{\ell} \sim \log n$ if $p=1$ \cite{ball}. 

The central result obtained by Watts and Strogatz \cite{watts} is very simple: even addition of a single shortcut to a large system decreases $\bar{\ell}$ noticeably, so that $\bar{\ell}$ is a quantity extremely sensitive to such "global" changes in a network as addition of a shortcut between far sites.

The main problem of the small network theory is to describe the crossover between the $p=0$ and $p=1$ regimes of a finite size network and to find a scaling form \cite{dg} for $\bar{\ell}(p,n)$. One should emphasize that in spite of a lot of effort to understand properties of small-world networks, the only exactly proved statement for $\bar{\ell}$ was the following: the scaling variable of the crossover under consideration is $pn$ -- e.g., see \cite{new1}. 
The reason of the failure is the difficulty of calculation of $\bar{\ell}$. In the present communication we consider very simple models with such a crossover between states with two different values of $\bar{\ell}$ for which those calculations may be made exactly. We shall present the scaling form of $\bar{\ell}$ and we shall describe $P(\ell)$ analytically. 
The answers are surprisingly close to the corresponding results obtained numerically for previously studied much more complex small-world networks.

Let us introduce the models. Let $n$ sites be arranged on a circle. Each of them is connected with two nearest neigbhours by links of unit length. We shall start our consideration from the most simple for consideration model with all the bonds on the circle oriented in one direction as shown in Fig. 1(a). Afterwards we shall study the model with all undirected links [see Fig. 1(b)]. We add a new central point to a network and connect it with any other site chosen with the probability $p$ by an undirected link of length equal to $1/2$. In fact, it means that we select some sites with the probability $p$ and then connect all of them by undirected unit length links.

Let us find the distribution $P(\ell)$ of the shortest paths between pairs of sites. 
The calculations are most simple for the model with directed bonds showed in Fig. 1(a).
In the present short communication, we use, maybe, the simplest way to obtain the distribution.
At first, we shall obtain the probability $P(\ell,k)$, i.e. the probability that the shortest path from one site to another is $\ell$ when the distance between them counted along the ring is $k$: $\sum_{\ell = 1}^{k} P(\ell , k) = 1$.
We shall find $P(\ell \leq k,k)$ for several small values of $k$ and shall demonstrate that it can easily be described by a general formula for all $\ell \leq k$ and $k$.

To find $P(\ell,k)$ for the model shown in Fig. 1(a), one has to consider all possible configurations of links connecting the center and sites $i=0,1,2, \ldots, k$ and to calculate statistical weights for the shortest path to be equal to a particular value $\ell$. In such a way, one may easily obtain the following expressions for small $\ell$ and $k$:
  
\begin{eqnarray}
\label{e1}
P(1,1) & = & 1 ,  \nonumber \\[5pt]
P(1,2) & = & p^2 , \nonumber \\
P(2,2) & = & 1 - p^2 , \nonumber \\[5pt]
P(1,3) & = & p^2 , \nonumber \\
P(2,3) & = & 2 p^2 (1-p) , \nonumber \\
P(3,3) & = & 1 - p^2 \cdot 1 - 2p^2 (1-p) , \nonumber \\[5pt]
P(1,4) & = & 1 \cdot p^2 (1-p)^0 , \nonumber \\
P(2,4) & = & 2 p^2 (1-p)^1 , \nonumber \\
P(3,4) & = & 3 p^2 (1-p)^2 , \nonumber \\
P(4,4) & = & 1 - p^2 [1 \cdot(1-p)^0 + 2(1-p)^1 + 3(1-p)^2] , \nonumber \\[5pt]
\ldots\ldots
\end{eqnarray}
The next expressions for $P(\ell,k)$ show the same trend. Thus, 
the general form of $P(\ell,k)$ is 
\begin{eqnarray}
\label{e2}
P(\ell<k,k) & = & \ell p^2 (1-p)^{\ell-1} , \nonumber \\
P(\ell=k,k) & = & 1- p^2 \sum_{i=0}^{k-1} i \,(1-p)^{i-1}.
\end{eqnarray} 
The distribution of shortest distances between pairs of sites for  the model under consideration looks as the following:
\begin{equation}
\label{e3}
P(\ell) = \frac{1}{n-1} \sum_{k=1}^{n-1} P(\ell,k) = \frac{1}{n-1} \sum_{k=\ell}^{n-1} P(\ell,k) . 
\end{equation}
Inserting Eqs. (2) into Eq. (3) we get immediately 
\begin{equation}
\label{e4}
P(\ell) = \frac{1}{n-1}\left[1 + (\ell-1)p + \ell(n-1-l)p^{2} \right] \; (1-p)^{\ell-1}  .
\end{equation}
If $p \to 0$, $P(\ell) \to 1/(n-1)$ for any $\ell \leq n-1$. If $p \to 1$, $P(\ell) \to \delta_{\ell,1}$, where $\delta_{i,k}$ is the Kronecker symbol.

The average shortest distance, $\bar{\ell}$ may be written as 
\begin{equation}
\label{e5}
\bar{\ell}= \sum_{\ell=1}^{n-1}\ell P(\ell) ,
\end{equation}
so using the previous distribution one gets
\begin{equation}
\label{e6}
\bar{\ell}= \frac{1}{n-1}\left[\frac{2-p}{p}n - \frac{3}{p^2}+ \frac{2}{p} + \frac{(1-p)^n}{p}\left(n-2+\frac{3}{p}\right)\right] .
\end{equation}
Thus, $\bar{\ell}(p \to 0) \to n/2$ and $\bar{\ell}(p \to 1) \to 1$.
The maximum value of the distribution, $P(\ell)$ is at the following point,
\begin{eqnarray}
\label{e7}
\ell_{max} = -\frac{1}{\log(1-p)} + \frac{1}{2p} + \frac{n-1}{2} \nonumber \\[2pt]
+ \sqrt{\frac{1}{\log^2(1-p)} + \frac{5}{4p^2} + \frac{n-3}{2p} + \frac{(n-1)^2}{4}}  .
\end{eqnarray}
Again, $\ell_{max}(p \to 0) \to n/2$ and $\ell_{max}(p \to 1) \to 1$. 

The Eqs. (4) and (6) are our main results for the defined model. To obtain a scaling description of the crossover region, one has to pass to the limits $n \to \infty$ and $p \to 0$ while the quantities 
$\rho \equiv p n$ and $z \equiv \ell/n$ are fixed.
In this limit, from Eq. (4) one obtains a continuous distribution $Q(z, \rho)$ of $z$:
\begin{equation}
\label{e8}
nP(\ell) \equiv Q(z, \rho) = [1 + \rho z + \rho^2 z(1-z)] \; e^{-\rho z} ,    
\end{equation}
$0 \leq z \leq 1$. 
In Fig. (2) we present $Q(z, \rho)$ for several values of $\rho$ as a function of shortest distance $z$.
The average value of the shortest distance may be found from Eq. (6) or (8):
\begin{equation}
\label{e9}
\frac{\bar{\ell}}{n} \equiv \bar{z} = \frac{1}{\rho^2} \; [2\rho - 3 + (\rho+3) \; e^{-\rho}] .
\end{equation}
The corresponding scaling function for $\ell_{max}$ is
\begin{equation}
\label{e10}
\frac{\ell_{max}}{n} \equiv z_{max} = \frac{1}{2\rho} \; \left[3 + \rho - \sqrt{9 + 2\rho + \rho^2}\right]  .
\end{equation}
The asymptotic behavior of $\bar{z}$ is $1/2-\rho^{2}/24$ ($\rho \to 0$) and $2/\rho$ ($\rho \gg 1$) and of $z_{max}$ is $1/3-2\rho/27$ ($\rho \to 0$) and $1/\rho$ ($\rho \gg 1$).
From Eqs. (\ref{e8}) and (\ref{e9}), one sees that $\overline{(z-\bar{z})^2} \simeq 2/\rho^2$ when $\rho \gg 1$, so $\sqrt{\,\overline{(z-\bar{z})^2}}/\bar{z} \to 1/\sqrt{2}$ in that limit.

In Figure 3, we plot $\bar{\ell}$ and $\ell_{max}$ vs $\rho$. Note, that the plot of $\bar{\ell}/n$ resembles strikingly the results obtained numerically (e.g., compare with \cite{watts,new1,barr2}) and  by series expansions \cite{new2} for ordinary small-world networks, although we consider much more simple network. The analytical result for the distribution $Q(z)$ presented in Fig. 2 looks very similar to the corresponding numerical one obtained by Barrat and Weight \cite{barr2} for a standard small-world network.

The model we have studied above looks in some sense rather artificial because of the directed links on the circle. Now we shall consider a more naturally looking model without directed links [see Fig. 1(b)]. We can proceed in the same way as before to obtain the following probabilities for small $k$ (although, the procedure is more tedious in this case, since now to obtain $P(\ell,k)$ one has to count over all configurations of links connecting the center and sites $-k+2,-k+3, \ldots, 2k-3,2k-2$),
\begin{eqnarray}
\label{e11}
P(1,1) & = & 1 ,  \nonumber \\[5pt]
P(1,2) & = & p^2 , \nonumber \\
P(2,2) & = & 1 - p^2 , \nonumber \\[5pt]
P(1,3) & = & p^2 , \nonumber \\
P(2,3) & = & p^2 (2-p)(2-2p) , \nonumber \\
P(3,3) & = & (1-p)^2 [1 + 2p - 2p^2] , \nonumber \\[5pt]
P(1,4) & = & p^2 , \nonumber \\
P(2,4) & = & p^2 (2-p)(2-2p) , \nonumber \\
P(3,4) & = & p^2 (1-p)^2 (2-p)(4-3p) , \nonumber \\
P(4,4) & = & (1-p)^4 (1 + 4p - 3p^2) , \nonumber \\[5pt]
P(1,5) & = & p^2 , \nonumber \\
P(2,5) & = & p^2 (2-p)(2-2p) , \nonumber \\
P(3,5) & = & p^2 (1-p)^2 (2-p)(4-3p) , \nonumber \\
P(4,5) & = & p^2 (1-p)^4 (2-p)(6-4p) , \nonumber \\
P(5,5) & = & (1-p)^6 (1 + 6p - 4p^2) , \nonumber \\[5pt]
\ldots\ldots
\end{eqnarray}
(One may check that Eqs. (\ref{e11}) hold for any possible $n$.) 
As before, these relations may be written in the following general form,
\begin{eqnarray}
\label{e12}
P(1,1) & = & 1 , \nonumber \\
P(\ell=1,k) & = & p^2 , \nonumber \\
P(2 \leq l<k,k) & = & p^2 (1-p)^{2l-4}(2-p)(2l-2-\ell p) , \nonumber \\
P(\ell=k,k) & = & (1-p)^{2k-4}[1 + (2k-4)p - (k-1)p^2]  \nonumber \\ 
         & = & 1 - \sum_{\ell=1}^{k-1} P(\ell,k) .
\end{eqnarray}  
For an odd $n$, for instance, one can write 
\begin{equation}
\label{e13}
P(\ell) =\frac{2}{n-1}\sum_{k=1}^{(n-1)/2}P(\ell,k) .
\end{equation}
Substituting Eq. (\ref{e12}) into Eq. (\ref{e13}) one gets
\begin{equation}
\label{e14}
P(1) = \frac{2}{n-1}\left(1 + \frac{n-3}{2}p^2 \right)
\end{equation}
and
\begin{eqnarray}
\label{e15}
P(\ell \geq 2) = [2 + 4(\ell-2)p +2(\ell-1)(2n-4\ell-3)p^2 - \nonumber \\ 
2(2\ell-1)(n-2\ell-1)p^3 + \ell(n-2\ell-1)p^4 ]\frac{(1-p)^{2\ell-4}}{n-1} .\nonumber 
\end{eqnarray}

This expression for $P(\ell)$ looks slightly different from Eq. (\ref{e4}) for the previous model. But when we take the scaling limits, we find the continuous distribution 
$nP(\ell) \equiv Q^{\prime}(z,\rho)=2Q(2z,\rho)$, which differs from the corresponding distribution $Q(z,\rho)$ for the model with oriented links [see Eq. (\ref{e8})] only by a scaling factor of two: $z  \to 2z$ (now $0 \leq z \leq 1/2$). Therefore, after that trivial change of a scale, Eqs. (\ref{e9}) and (\ref{e10}) and plots in Fig. 3 are also valid in the present case. In fact, the results for both models under consideration turns to be the same.

In summary, we have studied two networks which are simple analogies of small-world networks. We have obtained explicit expressions for the distribution $P(\ell)$ of the shortest paths between pairs of sites and for the average shortest path $\bar{\ell}$. The scaling functions look similar to the corresponding ones obtained numerically for more complex small-world networks \cite{watts,new1,barr2}. The studied networks (especially, the second one) model a rather real for our world situation in which all far connections occur through some common center.

Note that we have used here the simplest but not the most rigorous way to find an exact solution of the models under consideration. In fact, we have demonstrated above how to guess the solution, so the problem of searching for explicit solutions for small-world networks remains open. 

One should notice finally, that the above considered models differ from the original small-world networks \cite{watts,new1} in two aspects. One of them is evident: the studied models are of mean-field nature since all the selected sites are connected together by shortcuts. Another aspect is that, although, the original small-world networks are in an ordered state for $p=0$ and they are random graphs for $p=1$, in our case, the networks are ordered for both $p=0,1$.
Nevertheless, as we have shown, the results for our simple models resemble strikingly the corresponding answers for the original small-world networks. What is the reason for such a similarity? First of all, we have to say that the obtained crossover behavior seems to be independent of the mean-field nature of our models. Second, the scaling regime is observed in the limit $p \to 0$ when all such networks are in the disordered state. Nevertheless, we have solved analytically the model in which the links connected with the central site in Fig. 1 are directed chaotically 
\cite{dm}, so the crossover is between an ordered and disorded states. The obtained results for $\bar{\ell}$ again look similar to Fig. 3.

SND thanks PRAXIS XXI (Portugal) for a research grant PRAXIS XXI/BCC/16418/98. JFFM was partially supported by the projects PRAXIS/2/2.1/FIS/299/94, PRAXIS/2/2.1/FIS/302/94 and NATO grant No. CRG-970332. We also thank E.J.S. Lage for a thorough reading of the manuscript and A.V. Goltsev and A.N. Samukhin for many useful discussions.\\
$^{\ast}$      Electronic address: sdorogov@fc.up.pt\\
$^{\dagger}$   Electronic address: jfmendes@fc.up.pt

\end{multicols}

\begin{figure}
\epsfxsize=85mm
\epsffile{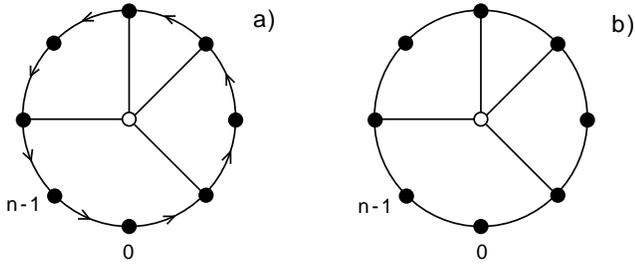}
\caption{Models under consideration. $n$ sites on the circle are connected by unit length links. A central site is connected to a particular site chosen with the probability $p$ by a link of half length. The links connected with the center are not directed. a) All links on the circle have a defined orientation. b) There are no directed links in the network.}
\end{figure}

\begin{figure}
\epsfxsize=90mm
\epsffile{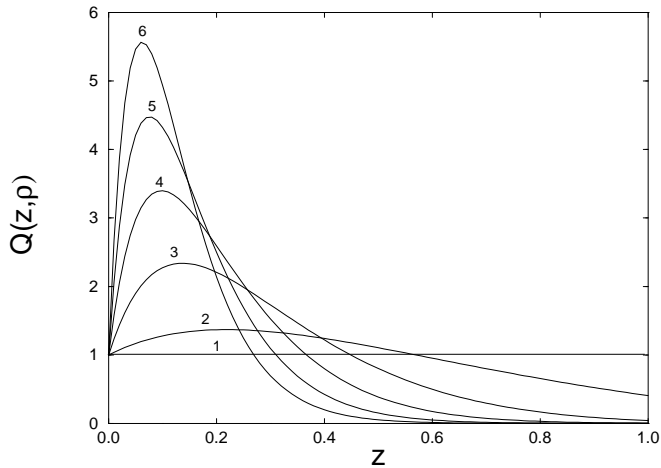}
\caption{The distribution $Q(z,\rho)$ of normalized shortest paths $z \equiv \ell/n$ between pairs of sites for different values of $\rho \equiv pn$ for the model shown in Fig.1 (a) [see Eq. (\protect\ref{e8})]. Curves labeled by numbers from 1 to 6 correspond to $\rho = 0$, $2$, $5$, $8$, $11$ and $14$. For the model without directed links, the corresponding distribution equals $2 Q(2z,\rho)$.
}
\end{figure}

\begin{figure}
\epsfxsize=90mm
\epsffile{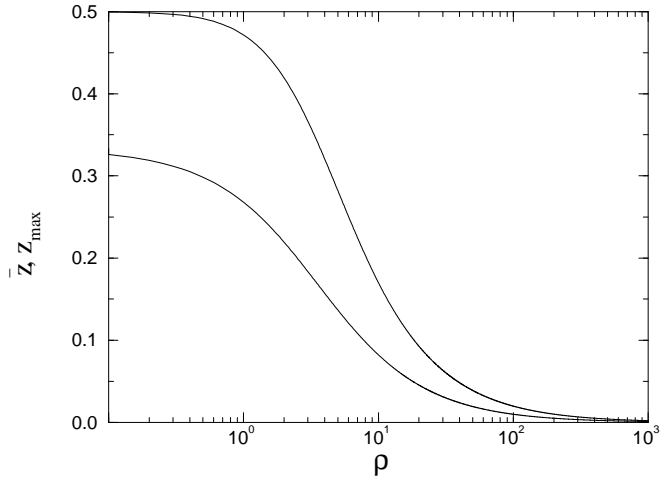}
\caption{The scaling functions $\bar{z} \equiv \bar{\ell}/n$ and $z_{max} \equiv \ell_{max}/n$ vs. $\rho \equiv pn$ for the model with directed links. The upper curve corresponds to $\bar{z}$.
For the model without directed links, one should change $\bar{z} \to 2 \bar{z}$ and $z_{max} \to 2 z_{max}$.
}
\end{figure}

\end{document}